\documentclass[
 ,draft            % use draft while you are working on the paper
  ]
  {aipproc}

\layoutstyle{6x9}

\newcommand{\includefigurev}[3]{}
   \newcommand{\includefigureh}[3]{}
\newcommand{\includefigurevh}[4]{}
\newcommand{\includefigure}[2]{}

\newcommand{\epsfboxm}[1]{}
\def\bit{\begin{itemize}}                                                      
\def\eit{\end{itemize}}
 \newcommand{\note}[1]{}
\newcommand{\delhad}{\mbox{$\Delta \alpha_{\rm had}^{(5)}(M_Z)$}} 
                   
\newcommand{\msb}{\mbox{$\overline{\rm{MS}}\ $}}                                
\newcommand{\mt}{\mbox{$m_t$}}                                                  
\newcommand{\mh}{\mbox{$M_H$}}                                                  
\newcommand{\mz}{\mbox{$M_Z$}}                                                  
\newcommand{\mw}{\mbox{$M_W$}}                                                  
                                      
\newcommand{\als}{\mbox{$\alpha_s$}}                                            
\newcommand{\suf}{\mbox{$SU(5)\ $}}

\newcommand{\skipblk}[1]{}                                                      
\def\bqa{\begin{eqnarray}}                                                      
\def\eqa{\end{eqnarray}}                                                        
\newcommand{\ee}{\mbox{$e^+ e^-$}}

\newcommand{\etal}{{\em et al., }}

\newcommand{\sto}{\mbox{$SU(2) \x U(1)\ $}}                                       
                                               
\newcommand{\x}{\mbox{$\times$}}

\newcommand{\sinn}{\mbox{$\sin^2\theta_W\,$}}                                   
                                            
\newcommand{\snu}{\mbox{$\stackrel{(-)}{\nu}$}}                                 
\newcommand{\beq}{\begin{equation}}                                             
\newcommand{\eeq}{\end{equation}}

\newcommand{\RA}{\mbox{$\rightarrow$}}

\def\mxth{\mathsurround=0pt }
\def\xversim#1#2{\lower2.pt\vbox{\baselineskip0pt \lineskip-.5pt
  \ialign{$\mxth#1\hfil##\hfil$\crcr#2\crcr\sim\crcr}}}             
\def\simgr{\mathrel{\mathpalette\xversim >}}                                    
\def\simle{\mathrel{\mathpalette\xversim <}}

\begin{document}
\DeclareGraphicsExtensions{.jpg,.pdf,.mps,.png}

\title{Electroweak Physics}

\author{Paul Langacker}{
  address={Department of Physics and 
Astronomy, University of Pennsylvania \\
Philadelphia, PA 19104, USA }}

\begin{abstract}
The results of high precision weak neutral current
(WNC), $Z$-pole, and high energy collider electroweak experiments 
have been the primary prediction and test of electroweak unification.
The electroweak program is briefly reviewed from a historical perspective.
Current changes, anomalies, and things to watch are summarized,
and the implications for the standard model and beyond discussed.
\end{abstract}

\maketitle

\section{The $Z$, the $W$, and the Weak Neutral Current}

The weak neutral current was a critical prediction of the electroweak
standard model (SM)~\cite{sirlinfest,general}. Following its discovery in 1973
by the Gargamelle and HPW experiments, there
were generations of ever more precise WNC experiments, typically at the
few~\% level.  These included
pure weak $\nu N$ and $\nu e$ scattering processes, and
weak-electromagnetic interference processes such as polarized
$e^{\uparrow \downarrow}D$ or $\mu N$, $\ee \RA $ (hadron or charged lepton)
cross sections and asymmetries below the $Z$ pole, and parity-violating
effects in heavy atoms (APV).  There were also  early direct observations
of the $W$ and $Z$ by UA1 and UA2. The early 1990's witnessed the very precise
$Z$-pole experiments at LEP and the SLC, in which the lineshape, decay modes,
and various asymmetries were measured at the 0.1\% level. The subsequent
LEP 2 program at higher energies measured \mw, searched for the Higgs
and other new particles, and constrained anomalous gauge self-interactions.
Parallel efforts at the Tevatron by CDF and D\O \
led to the direct discovery of the $t$ and  measurements of \mt \ and \mw,
while a fourth generation of weak neutral current experiments
continued to search for new physics to which the (more precise) $Z$-pole
experiments were blind. 
The program
was supported by theoretical efforts in the
calculation of QCD and electroweak
radiative corrections; the expectations for observables
in the standard model,
large classes of extensions, and alternative models; and global
analyses of the data.

The precision program
has established
that the standard model (SM) is correct and unique to first approximation,
establishing the gauge
principle as well as the SM gauge group and representations;
shown that the SM is correct at loop level, confirming the basic principles of
renormalizable gauge theory and allowing the successful prediction 
or constraint on $m_t$, $\alpha_s$, and the Higgs mass $M_H$;
severely constrained new  physics at the TeV scale, with the
ideas of unification strongly favored over TeV-scale  compositeness; and
yielded precise values for the  gauge couplings, consistent with 
(supersymmetric) gauge unification.

\section{Results before the LEP/SLD era}

Even before the beginning of the $Z$-pole experiments at LEP and SLC in 1989,
the precision  program had established~\cite{general}-\cite{costa}:
\begin{itemize}
\item Global analyses of all data carried more information than
the analysis of individual experiments, but care has to be taken
with systematic and theoretical uncertainties.
\item The SM is correct to first approximation.
The four-fermion operators for $\nu q$, $\nu e$,
and $eq$ were uniquely determined,
in agreement with the standard model, in model (i.e., gauge group) independent 
analyses. 
The $W$ and $Z$ masses agreed with the expectations
of the \sto gauge group and canonical Higgs mechanism, eliminating contrived
alternative models with the
same four-fermi interactions as the standard model.
       \item QCD evolved structure functions and 
      electroweak radiative corrections  were necessary  for the agreement
of theory and experiment.
        \item The weak mixing angle (in the on-shell renormalization scheme) 
was determined to be \sinn = 0.230 $\pm 0.007$; consistency of the various
observations, including radiative corrections,  required
$m_t < 200$ GeV.
\item Theoretical uncertainties, especially in the $c$ threshold 
in deep inelastic weak  charge current (WCC) scattering,
dominated.
        \item The combination of WNC and WCC data uniquely
determined the $SU(2)$ representations of all of the known fermions,
i.e.,  $\nu_e$ and $\nu_\mu$, as well as the $L$ and
$R$ components of the $e, \ \mu, \ \tau, \ d, \, s, \, b, \ u,$ and $c$~\cite{unique}.
In particular,  the left-handed $b$ and $\tau$ were  the
lower components of $SU(2)$ doublets, implying unambiguously that the $t$ quark
and $\nu_\tau$ had to exist.
This was independent of theoretical arguments based on
anomaly cancellation (which could have been evaded in alternative models
involving a vector-like third family), and of
constraints on \mt \ from electroweak loops.
        \item The electroweak gauge couplings were
well-determined, allowing a detailed comparison with the gauge
unification predictions of the simplest grand unified theories (GUT).
Ordinary
    \suf was excluded (consistent with the non-observation of proton decay),
but the supersymmetric extension was allowed, ``perhaps even the first harbinger of supersymmetry''~\cite{amaldi}.
%i.e., that the data
%was ``consistent with SUSY
%GUTS and perhaps even the first harbinger of supersymmetry''~\cite{amaldi}.
        \item There were stringent limits on new physics at the TeV scale, including
additional $Z'$ bosons, exotic fermions (for which both WNC and WCC 
constraints were crucial), exotic Higgs representations,
 leptoquarks, and new four-fermion operators.
\end{itemize}

%\includefigurevh{1}{xxnq_new.pdf}{-1cm}{-4cm}
%\includefigureh{1}{xxnue.pdf}{3.5cm}
%****************************************************************************
%%******************************************************************************
%******************************************************************************

\section{The LEP/SLC Era}
The LEP/SLC era greatly improved the precision of the electroweak program.
It allowed the differentiation between non-decoupling extensions to the
SM (such as most forms of dynamical symmetry breaking and other types
of TeV-scale compositeness), which typically predicted several
\% deviations, and decoupling extensions (such as most of the 
parameter space for supersymmetry), for which the deviations are
typically 0.1\%.

The first phase of the LEP/SLC program involved running at the $Z$
pole, $e^+ e^- \rightarrow Z \rightarrow \ell^+ \ell^-, \ \
  q \bar{q},$ and $\nu \bar{\nu}$. During the period 1989-1995 the
four LEP experiments ALEPH, DELPHI, L3, and OPAL at CERN
observed $\sim  2 \times 10^{7} Z's$. The SLD experiment at the SLC at
SLAC observed some $5 \times 10^5$ events. Despite the much lower statistics,
the SLC had the considerable advantage of a highly polarized $e^-$ beam,
with $P_{e^-} \sim$ 75\%. There were quite a few $Z$ pole observables,
including:
 \begin{itemize}
   \item The lineshape: $M_Z, \Gamma_Z,$ and the peak cross section $ \sigma$.
   \item The branching ratios for $e^+e^-,\ \mu^+ \mu^-,\ \tau^+ \tau^-, 
\ q \bar{q},\ c \bar{c},\ b \bar{b},$ and $ s \bar{s}$. One could also determine
the invisible width, $\Gamma({\rm inv})$, from which
one can derive the number 
%$\nu \bar{\nu} \Rightarrow 
$N_\nu = 2.986
\pm 0.007$ of active (weak doublet) neutrinos with 
       $m_\nu < M_Z/2$, i.e., there are only 3 conventional families with 
light neutrinos. $\Gamma({\rm inv})$ also constrains other invisible
particles, such as light sneutrinos and the light majorons associated with some 
models of neutrino mass.
   \item A number of asymmetries, including forward-backward (FB) asymmetries; 
the $\tau$ polarization, $P_\tau$;  the polarization asymmetry $A_{LR}$ associated
with $P_{e^-}$; and
mixed polarization-FB asymmetries.
    \end{itemize}
The expressions for the observables are summarized in~\cite{sirlinfest,general},
and the experimental values and SM predictions in
Table~\ref{tab1}.
The precision of the $Z$ mass determination was extraordinary for a high
energy experiment.
These combinations of observables could be used to isolate many
$Z$-fermion couplings, verify lepton family universality,
determine \sinn in numerous ways, and determine or constrain \mt, \als, and \mh.
 LEP and SLC simultaneously carried out other  programs,
most notably studies and tests of QCD, and heavy quark physics.

\small

\begin{table} \centering
\begin{tabular}{|l|c|c|c|r|}
\hline Quantity & Group(s) & Value & Standard Model & pull \\ 
\hline
$M_Z$ \hspace{14pt}      [GeV]&     LEP     &$ 91.1876 \pm 0.0021 $&$ 91.1874 \pm 0.0021 $&$ 0.1$ \\
$\Gamma_Z$ \hspace{17pt} [GeV]&     LEP     &$  2.4952 \pm 0.0023 $&$  2.4972 \pm 0.0011 $&$-0.9$ \\
{ $\Gamma({\rm had})$\hspace{8pt}[GeV]}&  
LEP  &$  1.7444 \pm 0.0020 $&$  1.7436 \pm 0.0011 $&  ---  \\
{ $\Gamma({\rm inv})$\hspace{11pt}[MeV]}& LEP  &$499.0    \pm 1.5    $&$501.74   \pm 0.15   $&  ---  \\
{ $\Gamma({\ell^+\ell^-})$ [MeV]}&     LEP     &$ 83.984  \pm 0.086  $&$ 84.015  \pm 0.027  $&  ---  \\
$\sigma_{\rm had}$ \hspace{12pt}[nb]&LEP    &$ 41.541  \pm 0.037  $&$ 41.470  \pm 0.010  $&{ $1.9$} \\
$R_e$                         &     LEP     &$ 20.804  \pm 0.050  $&$ 20.753  \pm 0.012  $&$ 1.0$ \\
$R_\mu$                       &     LEP     &$ 20.785  \pm 0.033  $&$ 20.753  \pm 0.012  $&$ 1.0$ \\
$R_\tau$                      &     LEP     &$ 20.764  \pm 0.045  $&$ 20.799  \pm 0.012  $&$-0.8$ \\
\hline
$A_{FB} (e)$                  &     LEP     &$  0.0145 \pm 0.0025 $&$  0.01639\pm 0.00026$&$-0.8$ \\
$A_{FB} (\mu)$                &     LEP     &$  0.0169 \pm 0.0013 $&$                    $&$ 0.4$ \\
$A_{FB} (\tau)$               &     LEP     &$  0.0188 \pm 0.0017 $&$                    $&$ 1.4$ \\
$R_b$                         &  LEP/SLD  &$  0.21664\pm 0.00065$&$  0.21572\pm 0.00015$&$ 1.1$ \\
$R_c$                         &  LEP/SLD  &$  0.1718 \pm 0.0031 $&$  0.17231\pm 0.00006$&$-0.2$ \\
$R_{s,d}/R_{(d+u+s)}$         &     OPAL    &$  0.371  \pm 0.023  $&$  0.35918\pm 0.00004$&$ 0.5$ \\
$A_{FB} (b)$                  &     LEP     &$  0.0995 \pm 0.0017 $&$  0.1036 \pm 0.0008 $&{ $-2.4$} \\
$A_{FB} (c)$                  &     LEP     &$  0.0713 \pm 0.0036 $&$  0.0741 \pm 0.0007 $&$-0.8$ \\
$A_{FB} (s)$                  &DELPHI/OPAL&$  0.0976 \pm 0.0114 $&$  0.1037 \pm 0.0008 $&$-0.5$ \\
$A_b$                         &     SLD     &$  0.922  \pm 0.020  $&$  0.93476\pm 0.00012$&$-0.6$ \\
$A_c$                         &     SLD     &$  0.670  \pm 0.026  $&$  0.6681 \pm 0.0005 $&$ 0.1$ \\
$A_s$                         &     SLD     &$  0.895  \pm 0.091  $&$  0.93571\pm 0.00010$&$-0.4$ \\
\hline
$A_{LR}$ (hadrons)            &     SLD     &$  0.15138\pm 0.00216$&$  0.1478 \pm 0.0012 $&{ $1.7$} \\
$A_{LR}$ (leptons)            &     SLD     &$  0.1544 \pm 0.0060 $&$                    $&$ 1.1$ \\
$A_\mu$                       &     SLD     &$  0.142  \pm 0.015  $&$                    $&$-0.4$ \\
$A_\tau$                      &     SLD     &$  0.136  \pm 0.015  $&$                    $&$-0.8$ \\
$A_e (Q_{LR})$                &     SLD     &$  0.162  \pm 0.043  $&$                    $&$ 0.3$ \\
$A_\tau ({\cal P}_\tau)$      &     LEP     &$  0.1439 \pm 0.0043 $&$                    $&$-0.9$ \\
$A_e ({\cal P}_\tau)$         &     LEP     &$  0.1498 \pm 0.0048 $&$                    $&$ 0.4$ \\
$Q_{FB}$                      &     LEP     &$  0.0403 \pm 0.0026 $&$  0.0424 \pm 0.0003 $&$-0.8$ \\
\hline
\end{tabular}
\caption{Principal $Z$-pole observables, their experimental values, 
theoretical predictions using the SM parameters from the global best
fit as of 1/03 (updated from~\cite{general}), and pull
(difference from the prediction divided by the uncertainty).
See~\cite{sirlinfest} for definitions of the quantitites.
$\Gamma({\rm had})$, $\Gamma({\rm inv})$, and $\Gamma({\ell^+\ell^-})$ are not 
independent.}
%independent, but are included for completeness.}
\label{tab1}
\end{table}

\normalsize

LEP~2 ran from 1995-2000, with energies gradually increasing from $\sim 140$ to $\sim 209$ GeV.
The principal electroweak results were precise measurements of the $W$ mass, as well
as its width and branching ratios (these were measured independently at the Tevatron);
a measurement of  $e^+ e^- \RA W^+ W^-$,  $ZZ$, and single $W$,
as a function of center of mass (CM)
energy, which tests the cancellations between diagrams that is characteristic
of a renormalizable gauge field theory, or, equivalently, probes the triple
gauge vertices;
limits on anomalous quartic gauge vertices;
measurements of various cross sections and asymmetries for
$e^+ e^- \RA f \bar{f}$ for $f=\mu^-,\tau^-,q,b$ and $c$, in reasonable
agreement with SM predictions;
a stringent lower limit of 114.4 GeV on the Higgs mass, and even hints
of an observation at $\sim$ 116 GeV;
and searches for supersymmetric or other exotic particles.

In parallel with the LEP/SLC program, there were 
precise ($< $ 1\%) measurements of atomic parity violation (APV) in cesium at Boulder,
along with the atomic calculations and related measurements needed for the
interpretation; precise new measurements of deep inelastic
scattering by the NuTeV collaboration at Fermilab, with
a sign-selected beam which allowed them to minimize the effects of the $c$ threshold
and reduce uncertainties to around 1\%; and few \% measurements of $\snu_\mu e$ by CHARM II
at CERN. Although the precision of these WNC processes was  lower
than the $Z$ pole measurements, they are still of considerable importance:
the $Z$ pole  experiments are blind to  types of new physics
that do not directly affect the $Z$,
such as a heavy $Z'$ if there is no $Z-Z'$ mixing,  while the WNC experiments are often very
sensitive. During the same period there were important electroweak results 
from CDF and D$\not{\! 0}$ at the Tevatron, most notably a precise value for $M_W$,
competitive with and complementary to the LEP~2 value; a direct measure of \mt,
and direct searches for  $Z'$, $W'$, exotic fermions, and supersymmetric particles.
Many of these non-$Z$ pole results are summarized in
Table~\ref{tab2}.

%\setlength{\oddsidemargin}{-2.0cm}
%Tevatron, LEP 2, $\nu N$, APV (1/03)} 
%\footnotesize
%\scriptsize
\small
\begin{table} \centering
\begin{tabular}{|l|c|c|c|r|}
\hline Quantity & Group(s) & Value & Standard Model & pull \\ 
\hline
$m_t$\hspace{8pt}[GeV]&Tevatron &$ 174.3    \pm 5.1               $&$ 174.4    \pm 4.4    $&$ 0.0$ \\
$M_W$ [GeV]    &      LEP       &$  80.447  \pm 0.042             $&$  80.391  \pm 0.018  $&$ 1.3$ \\
$M_W$ [GeV]    & Tevatron /UA2 &$  80.454  \pm 0.059             $&$                     $&$ 1.1$ \\
\hline
$g_L^2$        &     NuTeV      &$   0.30005\pm 0.00137           $&$   0.30396\pm 0.00023$&{ $-2.9$} \\
$g_R^2$        &     NuTeV      &$   0.03076\pm 0.00110           $&$   0.03005\pm 0.00004$&$ 0.6$ \\
$R^\nu$        &     CCFR       &$   0.5820 \pm 0.0027 \pm 0.0031 $&$   0.5833 \pm 0.0004 $&$-0.3$ \\
$R^\nu$        &     CDHS       &$   0.3096 \pm 0.0033 \pm 0.0028 $&$   0.3092 \pm 0.0002 $&$ 0.1$ \\
$R^\nu$        &     CHARM      &$   0.3021 \pm 0.0031 \pm 0.0026 $&$                     $&{ $-1.7$} \\
$R^{\bar\nu}$  &     CDHS       &$   0.384  \pm 0.016  \pm 0.007  $&$   0.3862 \pm 0.0002 $&$-0.1$ \\
$R^{\bar\nu}$  &     CHARM      &$   0.403  \pm 0.014  \pm 0.007  $&$                     $&$ 1.0$ \\
$R^{\bar\nu}$  &     CDHS 1979  &$   0.365  \pm 0.015  \pm 0.007  $&$   0.3816 \pm 0.0002 $&$-1.0$ \\
\hline
$g_V^{\nu e}$  &     CHARM II   &$  -0.035  \pm 0.017             $&$  -0.0398 \pm 0.0003 $&  ---  \\
$g_V^{\nu e}$  &      all       &$  -0.041  \pm 0.015             $&$                     $&$-0.1$ \\
$g_A^{\nu e}$  &     CHARM II   &$  -0.503  \pm 0.017             $&$  -0.5065 \pm 0.0001 $&  ---  \\
$g_A^{\nu e}$  &      all       &$  -0.507  \pm 0.014             $&$                     $&$ 0.0$ \\
\hline
$Q_W({\rm Cs})$&     Boulder    &$ -72.69   \pm 0.44              $&$ -73.10   \pm 0.04   $&$ 0.8$ \\
$Q_W({\rm Tl})$&Oxford/Seattle&$-116.6    \pm 3.7               $&$-116.7    \pm 0.1    $&$ 0.0$ \\
\hline
$10^3$ $\frac{\Gamma (b\rightarrow s\gamma)}{\Gamma_{SL}}$ & 
BaBar/Belle/CLEO &$ 3.48^{+0.65}_{-0.54} $&$ 3.20 \pm 0.09 $&$ 0.5$ \\
\hline
$\tau_\tau$ [fs] & direct/${\cal B}_e/ {\cal B}_\mu$ &$ 290.96 \pm 0.59 \pm 5.66 $&$ 291.90 \pm 1.81 $&$-0.4$ \\
$10^4$ $\Delta\alpha^{(3)}_{\rm had}$ & $e^+e^-$/$\tau$ decays &$ 56.53 \pm 0.83 \pm 0.64 $&$ 57.52 \pm 1.31 $&$-0.9$ \\
$10^9$ $(a_\mu - {\alpha\over 2\pi})$ & BNL/CERN &$ 4510.64 \pm 0.79 \pm 0.51 $&$ 4508.30 \pm 0.33 $&{ $2.5$} \\
\hline
\end{tabular}
\caption{Non-$Z$-pole observables, 1/03. The SM values are  updated from~\cite{general}.}
\label{tab2}
\end{table}

\normalsize

The effort required
the calculation of the needed electromagnetic,
electroweak,  QCD, and mixed radiative corrections
to the predictions of the SM. Careful consideration of
the competing definitions of the renormalized \sinn
was needed. 
The principal theoretical uncertainty is the hadronic
contribution \delhad \ to the running of 
$\alpha$ from its precisely known value at low energies
to the $Z$-pole, where it is needed to compare
the $Z$ mass with the asymmetries and other observables.
The radiative corrections, renormalization schemes, and
running of $\alpha$ are further discussed in~\cite{sirlinfest,general}.
The LEP Electroweak Working Group (LEPEWWG)~\cite{LEPEWWG} 
combined the results of
the four LEP experiments, and also those of SLD and some WNC and Tevatron
results, taking proper account of
common systematic and theoretical uncertainties.
Much theoretical effort  also went into the development,
testing, and comparison of radiative corrections packages, and
into the study of how various classes of new
physics would modify the observables, and how they could
most efficiently be parametrized.

\section{New Inputs, Anomalies, Things to Watch}
The results in Tables \ref{tab1} and \ref{tab2} are from 1/03,
while the fit results to be presented in the next Section
are from June 2002, updated from~\cite{general}. Jens Erler and I are currently performing
a new analysis for the next edition of the {\it Review Of Particle Physics};
it is useful to list here some of the things that have or will change
or to watch for.

\begin{itemize}
\item  As of 3/03, the LEP 2 value for the $W$ mass, $80.412(42)$ GeV,
is smaller than the previous value of  $80.447(42)$ GeV (used in Table \ref{tab2})
due to a revised ALEPH analysis~\cite{LEPEWWG}. This is closer to the
SM best fit prediction of  $80.391(18)$ GeV  and will lead to a small
increase in the predicted \mh. The Tevatron (CDF, D\O) Run I/UA2  value of
 $80.454(59)$ GeV is also slightly high. A new Run II value is expected.

\item The direct lower limit on the SM Higgs mass from LEP 2 is  
$\mh > 114.4$ GeV (95\% cl). The hints for events around 116 GeV 
were weakened in the final analysis.

\item A more precise \mt\ from the Tevatron Run II is awaited. 
The preliminary CDF and D\O \  values still have large
uncertainties~\cite{azzi}. A new preliminary D\O  \ analysis of their Run I data
yields 180.1 $\pm$ 5.4 GeV~\cite{azzi},
about 1$\sigma$ above the previous combined value of $174.3 \pm 5.1$ GeV. 
This will again lead to an increase in the \mh \ prediction.

\item There is a new  estimate of \als \ from the $\tau$
lifetime~\cite{tauwidth}, which is quite precise though theory-error dominated,
yielding \als$(M_\tau) = 0.356^{+0.027}_{-0.021}$, corresponding
to \als$(M_Z) = 0.1221^{+0.0026}_{-0.0023}$.

\item $A_{FB}(b)$, the forward-backward asymmetry into $b$ quarks, has the
value $0.0995(17)$,  2.4$\sigma$ below the standard model global
fit value of 0.1036(8). However, the SLD value for the related quantity
$A_b= 0.922(20)$ %(see the appendices in~\cite{pl} for the definitions)
is only  $0.6\sigma$ below the expected 0.9348(1), and
the hadronic branching fraction $R_b=0.2166(7)$, which at one time
appeared anomalous, is now only 1.1$\sigma$ above the expectation 0.2157(2). 
If not just a statistical fluctuation or systematic problem, $A_{FB}(b)$ could
be a hint of new physics.  However, any such effect should not contribute
too much to  $R_b$. The deviation is only around 5\%, but if the
new physics involved a 
 radiative correction to the coefficient $\kappa$ \
of \sinn, the change would have to be around
25\%. Hence, the new physics would most likely be at the
tree level, mainly increasing the magnitude of the right-handed coupling
to the $b$. This could be due to a heavy $Z'$ boson
with non-universal couplings to the third family~\cite{zpr,fcnc};
or to the mixing of the $b_R$ with exotic quarks~\cite{general,wagner},
such as with an $SU(2)$ doublet involving a heavy $B_R$ quark and a charge
$-4/3$ partner~\cite{wagner}.
There is a strong correlation between $A_{FB}(b)$ and
the predicted Higgs mass $M_H$ in the global fits. It
has been emphasized~\cite{chanowitz} that if one eliminated $A_{FB}(b)$ from the
fit (e.g., because it is affected by new physics) then the $M_H$
prediction would be lower, with the central value well below the
lower limit from the direct searches at LEP 2.
One resolution, assuming $A_{FB}(b)$  is due to an experimental
problem or fluctuation, is to invoke a supersymmetric extension
of the standard model with light sneutrinos, sleptons, and possibly
gauginos~\cite{alt}, which modify the radiative corrections and allow
an acceptable \mh.

\item The NuTeV collaboration at Fermilab~\cite{Nutev} have
reported the results of their deep inelastic measurements of
$\frac{\snu_\mu N \RA \snu_\mu X}{\snu_\mu N \RA \mu^{\mp}X}$.
%using their sign-selected beam. 
They  greatly reduce the
uncertainty in the charm quark threshold in the charged current denominator
by taking appropriate combinations of $\nu_\mu$ and $\bar{\nu}_\mu$.
They find a value for the on-shell weak angle $s_W^2$ of
0.2277(16), which is 3.0$\sigma$ above the global fit value of 0.2228(4).
The corresponding values for the left and right handed neutral current
couplings~\cite{general} are $g_L^2 = 0.3001(14)$ and
$g_R^2 = 0.0308(11)$, which are respectively 2.9$\sigma$ below and
0.7$\sigma$ above the expected 0.3040(2) and 0.0300(0).
Possible standard model explanations include an unexpectedly large
violation of isospin in the quark sea~\cite{Nutev}; an
asymmetric strange sea~\cite{davidson}, though NuTeV's data seems to favor the wrong 
sign for this effect;  nuclear shadowing effects~\cite{shadow}; or next to leading
order QCD effects~\cite{davidson}. 

More exotic interpretations could
include a heavy $Z'$ boson~\cite{zpr,davidson}, although
the standard GUT-type $Z's$ do not significantly improve the fits,
suggesting the need for a $Z'$ with ``designer'' couplings. Mixing
of the $\nu_\mu$ with a heavy neutrino could  account for
the effect~\cite{pati,takeuchi}, and also  for the slightly
low value for the number of light neutrinos $N_\nu = 2.986(7)$
from the $Z$ line shape when $N_\nu$ is allowed to deviate from 3 (this
shows up as a slightly high hadronic peak cross section in the standard
model fit with $N_\nu=3$)~\cite{general,zpr}. This mixing would also
affect muon decay, leading to an apparent Fermi constant smaller than the
true value. This would be problematic for the other $Z$-pole 
observables, but could be compensated by a large negative $T$
parameter~\cite{takeuchi}. However, such mixings would also lead
to a lower value for $|V_{ud}|$, significantly aggravating the
universality problem discussed below.

\item The Brookhaven $g_{\mu}-2$ experiment has reported a precise
new value~\cite{gmin2} using positive muons, leading to a new world average
$a_\mu = 11659203(8) \x 10^{-10}$. Improvements in the statistical
error from negative muon runs are anticipated.
Using
the theoretical value quoted by the experimenters for the
hadronic vacuum polarization contribution $a^{\rm had}_\mu$,
there was a small
discrepancy, with 
%the experimental $a_\mu$ larger than the standard model
%expectation by 
$a_{\mu}(exp) - a_{\mu}(SM)=(26 \pm 11) \x 10^{-10}$, a 2.6$\sigma$ effect.
The value and uncertainty in $a^{\rm had}_\mu$ are still
controversial\footnote{There are
also uncertainties in the smaller hadronic light by light
diagram. An unfortunate sign error increased the apparent  discrepancy with
 experiment at an earlier stage, but this has now been corrected.}:
subsequent analyses based on $e^{+}e^{-}$ data~\cite{Davier,Hagiwara} found
a $3\sigma$ discrepancy, while an analysis using $\tau$ decay data~\cite{Davier}
found a smaller $1\sigma$ effect. Recently
the CDM-2 collaboration found a mistake in their theoretical code for the
$e^{+}e^{-}\RA e^{+}e^{-} $ cross section, used to determine the luminosity
in the hadronic cross section~\cite{cmd2}. This should lower the 
discrepancy from $e^{+}e^{-}$ data to around $2\sigma$, closer to the $\tau$
value. New data from KLOE is anticipated.

Because of the confused situation with the vacuum polarization,
it is hard to know how seriously to take the discrepancy.
Nevertheless, $a_\mu $ is more sensitive than the electron moment to most
types of new physics, so it is important.
One obvious candidate for a new physics explanation would be supersymmetry~\cite{wjm},
with relatively low masses for the relevant sparticles and high $\tan \beta$
(roughly, one requires an effective mass scale of $\tilde{m} \sim 55 \ {\rm GeV} \
\sqrt{\tan \beta}$). There is a correlation between
the theoretical uncertainty in the vacuum polarization and in the
hadronic contribution to the running of $\alpha$ to the $Z$ pole~\cite{corr},
leading to a slight reduction in the predicted Higgs mass when $a_\mu$
is included in the global fit assuming the standard model.

\item \delhad, the hadronic contribution to the running of $\alpha$ up to the $Z$-pole,
introduces the largest theoretical uncertainty into the precision program, in
particular to the relation between \mz \ and the \msb \  weak angle $\hat{s}^2_Z$
(extracted mainly from the asymmetries). The uncertainty is closely related to that in
$a^{\rm had}_\mu$. There has been much recent progress using improved QCD calculations 
for the high energy part and  more  precise $e^{+}e^{-}$ data from  BES and elsewhere for the
low energy part.

\item 
A few years ago there was an apparent  2.3$\sigma$ discrepancy
between the measured value of the effective  (parity-violating) weak
charge $Q_W(Cs)$ measured in cesium~\cite{apv}, and the expected value.
Cesium has a single electron outside a tightly bound core, 
so the atomic matrix elements could be reliably calculated, leading
(it was thought)
to a combined theoretical and experimental uncertainly of around 0.6\%.
However, it turns out that there are surprisingly large
(O(1\%)) radiative corrections, including  Breit (magnetic) interactions,
vacuum polarization, vertex, and self-energy corrections~\cite{apv0,apv4}.
After a somewhat confusing period, the situation has apparently stabilized,
with the current value~\cite{apv4}, 
$Q_W({\rm Cs})=  -72.84(46)$,  in excellent agreement
with the SM expectation,  $-73.10(4)$. (An earlier $ -72.69(44) $ is listed
in Table \ref{tab1}.)

\item 
The unitarity of the CKM matrix can be partially tested
by the universality prediction that 
$\Delta \equiv 1 - |V_{ud}|^2 - |V_{us}|^2 - |V_{ub}|^2 $
should vanish. In particular $|V_{ud}|$ can be determined
by the ratio of $G^V_\beta/G_\mu$, where $G^V_\beta$ and
$G_\mu$ are respectively the vector coupling in $\beta$ decay
and the $\mu$ decay constant. The most precise determination
of $|V_{ud}|$ is from superallowed $0^+ \RA 0^+$ transitions,
currently yielding $|V_{ud}| = 0.9740(5)$~\cite{superallowed}.
Combining with  the PDG values for $|V_{us}|$ from kaon and hyperon decays and
$|V_{ub}|$ from $b$ decays, this yields 
a 2.3$\sigma$ discrepancy $\Delta = 0.0032(14)$, 
%It is unlikely
%that the uncertainties in $|V_{us}|$ or $|V_{ub}|$ could be
%responsible, 
suggesting either the presence of unaccounted-for new
physics, or, possibly, effects from  higher order isospin violation such
as nuclear overlap corrections.
However, the latter have been
carefully studied, so the effect may be real.
This problem has been around
for some time, but until recently less precise determinations from
neutron decay were consistent with universality.
Recently, a more precise measurement of the neutron decay
asymmetry has been made by the PERKEO-II group at ILL~\cite{neutron}.
When combined with the accurately known neutron lifetime, this allowed
the new determination $|V_{ud}|= 0.9713(13)$, implying
$\Delta = 0.0083(28)$, i.e., a 3$\sigma$ violation of unitarity. Note,
however, that this value is only marginally consistent with the value
obtained from superallowed transitions.

Mixing of the $\nu_\mu$ with a heavy neutrino, suggested as a solution
of the NuTeV anomaly, would mean that $G_\mu$ is larger than the apparent value
and would aggravate this discrepancy. ($\nu_e$ mixing would affect
$G^V_\beta$ and $G_\mu$ in the same way and have no effect.)
However, a very small mixing of the $W$ boson with a heavy $W'$ coupling
to right handed currents, as in left-right symmetric models,
could easily account for the discrepancy for the appropriate sign for the
mixing~\cite{lr}, especially if the right-handed neutrinos are Majorana
and too heavy to be produced in the decays.

The situation has recently become more complicated, by the suggestion that
the culprit may be in the long accepted value of  $|V_{us}|$. 
The BNL E865 experiment has recently performed a high statistics
measurement of the $K^{+}_{e3}$ branching ratio~\cite{Sher}, obtaining
a result 2.3$\sigma$ higher than the old measurements. This would
be sufficient to account for the entire discrepancy, but must be confirmed by
new analyses  and measurements from KLOE, CMD-2 and NA48.

\item 
The LEP and SLC $Z$-pole experiments are the most precise tests of the
standard electroweak theory, but they are insensitive to any new
physics that doesn't affect the $Z$ or its couplings. Non-$Z$-pole
experiments are therefore extremely important, especially given
the possible NuTeV anomaly. In the near future we can expect
new results in polarized M\o ller scattering from SLAC~\cite{moller}, and
in the QWEAK polarized electron experiment at Jefferson Lab~\cite{qweak}.
%as well as improved inputs concerning \mt, \mw, and \mh from the Tevatron and LHC.
%In the long term, very precise values of such quantities from a linear collider
%would be very useful, as would a $Z$-pole option, which could repeat
%the entire $Z$ pole program with much higher precision~\cite{giga}.
 \item
 Although the $Z$-pole program has ended for the time being, there are
prospects for future programs using the Giga-$Z$ option at a  linear collider, which
might yield a factor $10^2$ more events. This would 
enormously improve the sensitivity~\cite{giga}, but would also require a large
theoretical effort to improve the radiative correction 
calculations.

\end{itemize}

\section{Fit Results (06/02)}
As of June, 2002, the result of the global fit was
%$  M_H = 86^{+49}_{-32} \mbox{ GeV}$; 
% $ m_t  =174.2  \pm 4.4  \mbox{ GeV}$; 
% $ \alpha_s = 0.1210 \pm 0.0018$; 
% $ \hat{\alpha}(M_Z)^{-1} = 127.922 \pm 0.020$; and
%  $ \hat{s}^2_Z = 0.23110 \pm 0.00015$, with a
%$\chi^2/{\rm d.o.f.}$ of $49.0/40 (15\%)$.
\bqa
           M_H &=& 86^{+49}_{-32} \mbox{ GeV}, \nonumber \\
           m_t &=& 174.2  \pm 4.4  \mbox{ GeV}, \nonumber  \\
      \alpha_s &=& 0.1210 \pm 0.0018, \nonumber \\
  \hat{\alpha}(M_Z)^{-1} & = & 127.922 \pm 0.020 \nonumber \\
   \hat{s}^2_Z &=& 0.23110 \pm 0.00015, \nonumber \\
%  \bar{s}^2_\ell &=& 0.23139 \pm 0.00015, \nonumber \\
%          s^2_W &=& 0.22277 \pm 0.00035 \nonumber \\
%          s^2_{M_Z} &=& 0.23105 \pm 0.00008 \nonumber \\
% \Delta \alpha_{\rm had}^{(5)}(M_Z) &=& 0.02778 \pm 0.00020 \nonumber
\chi^2/{\rm d.o.f.} & = & 49.0/40 (15\%) \label{results}
\eqa
The precision data alone yield 
$m_t = 174.0^{+9.9}_{-7.4}$ GeV from loop corrections, in impressive
agreement with the direct Tevatron value $174.3 \pm 5.1$.
The result  \als $=0.1210 \pm 0.0018$ for the strong coupling is somewhat
above the previous world average $\als = 
0.1172(20)$, which includes other determinations, most of which
are dominated by theoretical uncertainties~\cite{hinchliffe}. 
This 
is due in part to the inclusion of the new $\tau$ lifetime result~\cite{tauwidth}.
(Without it, one would obtain \als=$0.1200 \pm 0.0028$.) 
The $Z$-pole value is insensitive to oblique (propagator) new physics, but is
very sensitive to non-universal new physics, such as those which affect
the $Z b \bar{b}$  vertex.

The prediction for the Higgs mass from indirect data, 
\mh $= 86^{+49}_{-32}$ GeV, should be compared with the 
direct LEP 2 limit %~\cite{lephiggs} 
$\mh \simgr 114.4 (95\%)$ GeV.
The theoretical range in the standard model is
 115 GeV $\simle \mh \simle$ 750 GeV,
where the lower (upper) bound is from vacuum stability (triviality).
In the MSSM, one has
 $\mh \simle 130$ GeV, while \mh \ can be as high as 150 GeV in generalizations.
 Including the direct LEP 2 exclusion results, one finds
$\mh < 215$ GeV at 95\%. 
%The probability distribution for \mh,
%including both direct and indirect constraints and
%updated from the analysis in~\cite{erler}\footnote{See also
%the study in~\cite{degrassi}},
% is shown in Figure \ref{higgspdf}.
\mh \ enters the expressions for the radiative corrections logarithmally. It is fairly 
robust to many types of new physics, with some exceptions. In particular,
a much larger \mh \
 would be allowed for
 negative values for the $S$ parameter or positive values for $T$.
The predicted value would decrease if new physics accounted for
the value of $A_{FB}(b)$~\cite{chanowitz}.

%\begin{figure}[h]
%\centering
%\includegraphics*[scale=0.6]{higgs_pdf}
%\caption{Allowed regions in \mh \  vs \mt \ from precision data,
%compared with the direct exclusion limits from LEP 2. Courtesy of Jens Erler.}
%\label{higgspdf}
%\end{figure}

\section{Beyond the Standard Model}
The 
$\rho_0$ or $S$, $T,$ and $ U$ parameters describe the tree level effects of
Higgs triplets, or the loop effects on the $W$ and $Z$ propagators due
to such new physics as
nondegenerate fermions or scalars, or chiral families (expected,
for example,  in extended technicolor). 
%The current values are
%$S = -0.14 \pm 0.10 (-0.08)$,
%$T = -0.15 \pm 0.12 (+0.09)$, and
%$U = 0.32 \pm 0.12 (+0.01)$  (2.6$\sigma$ from zero)
The current values are:
 \bqa  S &=& -0.14 \pm 0.10 (-0.08)  \nonumber  \\
T &=& -0.15 \pm 0.12 (+0.09)  \nonumber \\
U &=& 0.32 \pm 0.12 (+0.01)  \ \ \  (2.6\sigma) %\nonumber
\eqa
for $M_H = 115.6 \ (300)$ GeV,
where these represent the effects of new physics only (the \mt \ and \mh \ effects
are treated separately).
Similarly,
 $\rho_0  \sim 1 + \alpha T = 0.9997^{+0.0011}_{-0.0008}$
for $M_H = 73^{+106}_{-34}$
GeV and $S=U=0$. If one constrains
$T = U =0$, then $S=0.10^{+0.12}_{-0.30}$. There is a strong negative $S-\mh$
correlation, so that the Higgs mass constraint is relaxed to $M_H < 570$
GeV at 95\%. 
For \mh \ fixed at 115.6 GeV, one finds
$S = -0.040(62)$, which implies that the number of ordinary plus
degenerate heavy families
is constrained to be
$N_{\rm fam} = 2.81 \pm 0.29$.  This
is
complementary to the lineshape constraint,  
$N_\nu = 2.986 \pm 0.007 $, which  only applies  to neutrinos less massive
than $M_Z/2$. One can also restrict   additional nondegenerate families by allowing
both $S$ and $T$ to be nonzero, yielding
$N_{\rm fam} = 2.79 \pm 0.43$  for $T = -0.01 \pm 0.11$.

In the 
decoupling limit of supersymmetry, in which the sparticles are heavier than
$ \simgr 200-300$ GeV, there is little effect on the precision
observables, other than that there is necessarily 
  a light SM-like Higgs, consistent with the data. There is little
improvement on the SM fit, and in fact one can somewhat constrain
the supersymmetry breaking parameters~\cite{susy}.

Heavy $Z'$ bosons are predicted by many 
grand unified  and string theories~\cite{general}. Limits on the $Z'$ mass
are model dependent, but are typically  around $M_{Z'} > 500-800 $ GeV 
from indirect constraints from WNC and  LEP~2 data, with comparable
limits from direct searches at the Tevatron. $Z$-pole data
severely constrains the $Z-Z'$ mixing, typically
 $|\theta_{Z-Z'}| < {\rm few} \times 10^{-3}$.
A heavy $Z'$ would have many other theoretical  and
experimental implications~\cite{heavyz}.

Precision data  constrains mixings between ordinary and exotic
fermions, large extra dimensions, 
new four-fermion operators, and leptoquark bosons~\cite{general}.

Gauge unification is predicted in GUTs and some string theories.
The simplest non-supersymmetric unification is excluded by
the precision data. For the MSSM, and assuming 
no new thresholds between 1 TeV and the unification scale, one
can use the precisely known $\alpha$ and $\hat{s}^2_Z$
to predict $\als = 0.130 \pm 0.010$ and a unification scale 
$M_G \sim 3 \times 10^{16}$ GeV~\cite{polonsky}. The \als \ uncertainties are 
mainly theoretical, from the TeV and GUT thresholds, etc.
\als \ is high compared to the experimental value, but barely consistent 
given the uncertainties. 
$M_G$ is reasonable for a GUT (and
is consistent with simple seesaw models of neutrino mass),
but is somewhat below the expectations $\sim 5 \times 10^{17}$ GeV of the simplest
perturbative heterotic string models. However, this is only a
10\% effect in the appropriate variable
$\ln M_G$. The new exotic particles often present in such models
(or higher Ka\v c-Moody levels) can easily shift the $\ln M_G$
and \als \ predictions significantly, so the problem is really
why the gauge unification works so well.
It is always possible that the apparent success is accidental
(cf., the discovery of Pluto).

\section{Conclusions}

The precision $Z$-pole, LEP~2, WNC, and Tevatron experiments have
successfully tested the SM at the 0.1\% level, including electroweak loops, thus
confirming the gauge principle,
SM   group, representations, and the
basic structure of renormalizable field theory.
The standard model parameters $\sin^2 \theta_W$, $m_t$, and $\alpha_s$
were precisely determined.
In fact, \mt \ was successfully predicted from its indirect loop effects prior
to the direct discovery at the Tevatron, while the indirect value of \als,
mainly from the $Z$-lineshape, agreed with more direct QCD determinations.
Similarly, \delhad \ and $ M_H$ were constrained.
The indirect (loop) effects implied $M_H \simle 215$ GeV, while direct
searches at LEP~2 yielded $M_H > 114.5 $ GeV, with a hint of a signal at 116 GeV.
This range is consistent with, but does not prove, 
the expectations of the supersymmetric
extension of the SM (MSSM), which predicts a light SM-like Higgs for much of
its parameter space. The agreement of the data with the SM imposes
a severe constraint on possible new physics at the TeV scale,
and points  towards decoupling theories (such as most versions of
supersymmetry and unification), which typically lead to 0.1\% effects,
rather than TeV-scale compositeness (e.g., dynamical symmetry breaking
or composite fermions), which usually imply  deviations of several \% (and often
large flavor changing neutral currents). 
Finally, the precisely measured gauge couplings were consistent with the
simplest form of grand unification if the SM is extended to the MSSM.

\begin{theacknowledgments}

It is a pleasure to thank my collaborators, especially Jens Erler, for
fruitful interactions. 
 This work was supported in part by a Department of Energy grant DOE-EY-76-02-3071.

 \end{theacknowledgments}

\end{document}